\documentclass[aps,prb,superscriptaddress,twocolumn,floatfix,showpacs,amsmath]{revtex4-1}
\usepackage{graphicx}
\usepackage{floatrow}
\usepackage{float}
\restylefloat{table}
\usepackage{placeins}
\usepackage{caption}
\usepackage{latexsym}
\usepackage{psfrag}
\usepackage{dcolumn}
\usepackage{blindtext}
\usepackage{amsmath, amssymb}
\usepackage{bbold}
\usepackage{subcaption}
\usepackage[labelformat=parens,labelsep=quad,skip=3pt]{caption}
\usepackage[%
  colorlinks=true,
  urlcolor=blue,
  linkcolor=blue,
  citecolor=blue
]{hyperref}
\begin{document}

\title{Comparison of band-fitting and Wannier-based model construction for WSe$_2$}
\author{ James Sifuna}
\affiliation{ Departamento de Ciencias de la Tierra y
	F\'{\i}sica de la Materia Condensada, Universidad de Cantabria,
	Cantabria Campus Internacional,
	Avenida de los Castros s/n, 39005 Santander, Spain}
\affiliation{ Department of Natural Science, 
	The Catholic University of Eastern Africa,
	62157 - 00200, Nairobi, Kenya}
\affiliation{ Materials Modeling Group, 
	Department of Physics and Space Sciences,
	The Technical University of Kenya,
	52428-00200, Nairobi, Kenya}  

\author{ Pablo Garc\'{\i}a-Fern\'andez}
\affiliation{ Departamento de Ciencias de la Tierra y
             F\'{\i}sica de la Materia Condensada, Universidad de Cantabria,
             Cantabria Campus Internacional,
              Avenida de los Castros s/n, 39005 Santander, Spain}   
\author{ George S. Manyali}  
\affiliation{Computational and Theoretical Physics Group (CTheP), 
Department of Physical Sciences,
	Kaimosi Friends University College,
	385-50309, Kaimosi, Kenya}        
\author{ George  Amolo}
\affiliation{ Materials Modeling Group, 
	Department of Physics and Space Sciences,
	The Technical University of Kenya,
	52428-00200, Nairobi, Kenya}
    
\author{ Javier Junquera }
\affiliation{ Departamento de Ciencias de la Tierra y
	F\'{\i}sica de la Materia Condensada, Universidad de Cantabria,
	Cantabria Campus Internacional,
	Avenida de los Castros s/n, 39005 Santander, Spain}
          
\date{\today}

\begin{abstract}
Transition metal dichalcogenide materials $MX_2 (M=Mo,W;X=S,Se)$ are being thoroughly studied due to their novel two-dimensional structure, that is associated with exceptional optical and transport properties. 
From a computational point of view, Density Functional Theory simulations perform very well in these systems and are an indispensable tool to predict and complement experimental results. However, due to the time and length scales where even the most efficient DFT implementations can reach today, this methodology suffers of stringent limitations to deal with finite temperature simulations or electron-lattice coupling when studying excitation states: the unit cells required to study, for instance, systems with thermal fluctuations or large polarons would require a large computational power.
%
Multi-scale techniques, like the recently proposed Second Principles Density Functional Theory, can go beyond these limitations but require the construction of tight-binding models for the systems under investigation. In this work, we compare two such methods to construct the bands of WSe$_2$. In particular, we compare the result of (i) Wannier-based model construction with (ii) the band fitting method of Liu \textit{et al.,}\cite{3band}  where the top of the valence band and the bottom of the conduction band are modeled by three bands symmetrized to have mainly Tungsten $d_{z^2}$, $d_{xy}$ and $d_{x^2-y^2}$ character. 
Our results emphasize the differences between these two approaches and how band-fitting model construction leads to an overestimation of the localization of the real-space basis in a tight-binding representation.   

\end{abstract}


\maketitle

\section{Introduction}
\label{sec:intro}
Monolayers based on group-VIB transition metal dichalcogenides have recently emerged as semiconducting alternatives to graphene in which their two-dimensionality brings out new semiconducting physics\cite{nat}. Layered semiconductors based on transition-metal dichalcogenides evolve from indirect bandgap in the bulk limit to direct bandgap in the quantum two-dimensional (2D) limit. This crossover is usually  achieved by peeling off a multilayer sample to a single layer\cite{intro}. Numerous studies\cite{intro,nat,opt} predict that these two-dimensional materials are semiconducting, possess a direct band gap in the visible frequency range and that the electrons have high mobility at room temperature\cite{rt,rty,rtyy}.
\par WSe$_{2}$ monolayer is regarded as a semiconductor analog of graphene, with both the conduction and valence band edges located at the two corners of the first Brillouin zone (BZ), i.e., K and -K points. First-principles methods based on the  Density Functional Theory (DFT) are able to predict the  ground state properties of WSe$_2$.\cite{luoprb,rocha} However, these methods still display significant restrictions.
First, they are usually applied at zero temperature.
Second, given that they can only deal with a relatively small number of atoms in the simulation box, and although the field is rapidly evolving\cite{feliciano}, 
they are not well prepared to deal with complex dynamics including electron-phonon and electron-electron interactions, key to simulate polaron charge transport or excitonic phenomena.
 Our capacity to interpret or predict experimental results in these fields would be significantly improved with the advent of computational methods that go beyond the time and length scales that DFT methods can reach today. In this manner we would be able to efficiently sample systems to account for effects of temperature and carry out simulations where disorder is treated beyond the highly ordered crystalline cells involved in DFT simulations.
\par In this era, multi-scale implementations strongly borrow from  semi-empirical methods such as tight-binding\cite{tb} and force field schemes\cite{ff}. However, few of these schemes have been able to retain first-principles-like accuracy,\cite{giese} and flexibility, being limited when treating situations in which the key interactions involve very small energy differences($\approx$ meV/atom) and great accuracy is needed. One of the main drawbacks of many first-principles-based models is their inability to carry out simulations including simultaneously electronic\cite{electronic}  and lattice\cite{lattice} degrees of freedom. For example, Heisenberg, tight-binding and force-field models can be parameterized from first-principles focusing, respectively, on magnetic, band and vibrational variables while all degrees of freedom outside the scope of the method are completely neglected. Trying to move beyond this setting, the novel second-principles DFT approach (SPDFT)\cite{pablo} is able to incorporate both atomic and electronic degrees of freedom, simultaneously and with high accuracy, as well as a modest computational cost. This method includes; ($i$) an accurate model potential that describes the lattice-dynamical properties of the system\cite{scaleup} and ($ii$) one-electron Hamiltonian including electrostatics and short range tight-binding like interactions corrected with electron-electron correlations.~\cite{pablo} In addition, this approach describes the relevant electronic degrees of freedom and electron-phonon interactions\cite{pablo}. It is worth noting that the second-principles accuracy can be systematically improved to essentially match that of a first-principles DFT calculation, if needed. While the method has some limitations, like the inability to predict bond creation or destruction, most of the transport or optical properties of materials can be precisely calculated simply taking into account a small enough deformation of the bonds around the equilibrium distance. 
 
\par In this paper, we have developed a three and a nine-band model for a monolayer of WSe$_2$ as a first step to create a full SPDFT model for the material. Our main goal here is to compare our tight-binding approach, based on Wannier functions, with  other tight-binding techniques\cite {tb} where an analytical symmetry-based Hamiltonian is constructed {\it a-priori} selecting interactions with a pre-determined number of neighbors followed by a parametrization to fit the DFT bands. In the case of MSe$_2$ (M = Mo, W) Liu \textit{et al.,}\cite{3band} developed such a model to describe the top of the valence and the bottom of the conduction band using idealized M-$d_{z^2}$, $d_{xy}$ and $d_{x^2-y^2}$ orbitals. Here, we show the limitations of this scheme when the interactions are compared to those with explicit basis functions that are calculated from {\it ab-initio}, i.e. not fitting the DFT bands, using a Wannier function scheme\cite{wanniercode} running on the {\sc siesta}\cite{siesta} DFT code.

\begin{figure}[H]
	\includegraphics[width=1\textwidth]{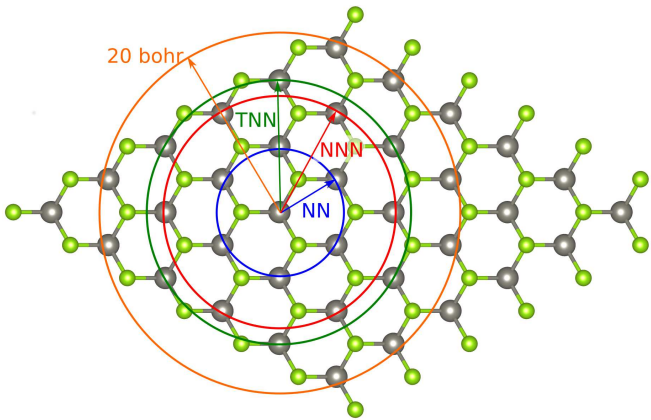}
	\caption{(Color online) Top view of monolayer WSe$_2$. The green balls represent selenium while the gray ball represent Tungsten. The blue circle denotes the Nearest Neighbor (NN) distances, the red curve, denotes the Next Nearest Neighbor (NNN) distances while the green curve denotes the Third Nearest Neighbor distances (TNN) as used in the TB model by Liu \textit{et al.}\cite{3band}. For comparison we show in orange the 20.0 Bohr distance limit.}
\label{fig:distances}		
\end{figure}
\section{Theory}
Just like in first-principles DFT, the goal of second-principles DFT is to describe the electrons in the system, and
the relevant electronic interactions in an efficient manner. Here we shall focus on the description of the valence and conduction states at the tight-binding level. The lattice of ionic cores comprised by the nuclei and the corresponding core electrons will not be modeled explicitly. 

\subsection{Parameter calculation}
\label{sec:parameters}
Textbooks tight-binding models are created by assuming the shape of the orbitals, for instance $s$, $p$ or $d$ functions centered around a particular atom, and then using symmetry\cite{tb} to calculate the degrees of freedom associated with the one-electron orbital-orbital interactions up to a particular range. In a second step the parameters associated with the degrees of freedom are determined by fitting to experimental or first-principles data. In the work by Liu \textit{et al.,}\cite{3band} this scheme was implemented to calculate the analytical Hamiltonian with interactions up to third neighbors (see Fig. \ref{fig:distances}) and the parameters where obtained by fitting to DFT bands.

Here we use an alternative way to construct a tight-binding-like Hamiltonian for monolayer WSe$_2$.
First of all, the electronic band structure is computed at the DFT level.
In this work, those calculations are carried out with the {\sc siesta} package.
Then, we proceed to the Wannierization of the top of the valence band and the
bottom of the conduction band manifold using the {\sc wannier90} code.
Finally, the one-electron Hamiltonian is written in the basis of the previously
obtained Wannier functions.
At the time of computing the Wannier functions, we can choose between
using ($i$) a maximally localization algorithm, 
where the spread of the localized
Wannier functions is minimized~\cite{metados}; 
or ($ii$) a ``maximally projected'' algorithm, where the projection 
on one of the atomic orbitals used in the basis set of {\sc siesta} is 
maximized.
The resulting Wannier orbitals, while expected to be larger than those obtained by maximum localization, retain the initial symmetry of the basis atomic orbital, a property that is very useful when building models. 
We would like to note here that, for the particular case of WSe$_{2}$ monolayers, the Wannier functions that we have obtained using both procedures are essentially the same so we have focused on using the second one which is faster and retains the symmetry of the basis function used to create the Wannier orbital. Finally, it is important to note that here the Wannier functions that act as the basis of the Hamiltonian are calculated explicitly, in contrast with the textbook procedure above where only their symmetry is used. 

In a final step, the Hamiltonian obtained from the wannierization procedure was trimmed for improved efficiency. In order to speed-up the calculation during a second-principles simulation we are interested in reducing the number of Hamiltonian elements to a minimum. The trimming procedure consists imposing a cutoff radius, $r_h$, to the Hamiltonian matrix elements between Wannier functions $a$ and $b$, $h_{ab}$. All matrix elements whose distance between the centroids $d_{ab}$ is larger than the cutoff radius are neglected. In a simulation we would select the cutoff radius as the smallest one where the comparison between DFT and second-principles bands is acceptable.

As discussed above, in both tight-binding model generating schemes we impose a maximum range for the interactions. The main question of the paper comes from the comparison of both methods when the range cutoff is the same, i.e. how well a method without explicit basis like the usual tight-binding compares with a formally correct explicit Wannier-based parameterization. 

\subsection{Technicalities}
\label{sec:fpdft}
We constructed our models following the discussion in the previous Section,
where the first-principles data were obtained from scalar-relativistic simulations of monolayer WSe$_2$ using the DFT formalism as implemented in the {\sc siesta} method\cite{siesta}. 
Exchange  and correlation were treated within the generalized gradient approximation using the PBE\cite{pbe} functional. 
Core electrons were replaced by {\it ab-initio} norm-conserving fully separable \cite{KB} Troullier-Martin pseudopotentials \cite{TroulMartin}. In {\sc{siesta}}, the one-electron eigenstates were expanded in a set of numerical atomic orbitals using the standard {\sc siesta} double-zeta polarized (DZP) basis.~\cite{DZP} 
A Fermi-Dirac distribution with a temperature of 8 meV was used to smear the occupancy of the one-particle electronic eigenstates that were calculated on a 20$\times$20$\times$1 Monkhorst and Pack grid.~\cite {monkhorst, soler_kpts}
Real space integration was carried over an uniform grid with an equivalent plane-wave cutoff of 600 Ry.
We optimized the geometry of the system introducing a vacuum of 20\AA ~between the WSe$_2$ layers to minimize the interactions. In this calculation, all the atomic coordinates were relaxed until the maximum component of the force on any atom was smaller than $0.01$ eV/\AA ~and the maximum component of the stress tensor was below 
0.0001 eV/\AA$^3$.
The band structure was calculated along the high-symmetry lines connecting the $\Gamma$ (0,0), K ($\frac{2}{3},\frac{1}{3}$) and M ($\frac{1}{2},\frac{1}{2}$) points in the 2D-Brillouin zone, as shown in Fig.~\ref{fig:3band} and 
Fig.~\ref{fig1}.

For the second-principles simulations we employed a methodology implemented in the {\sc scale-up} package\cite{pablo,scaleup,webscup}. 

The resulting potential yielded a qualitatively correct description of the band structure of WSe$_2$. 

\section{RESULTS AND DISCUSSION}
\label{sec:results}

\subsection{The Wannier orbitals}
\label{sec:wanniers}

\begin{figure}[H]
	\centering
	\begin{subfigure}[b]{0.3\textwidth}
		\includegraphics[width=\textwidth]{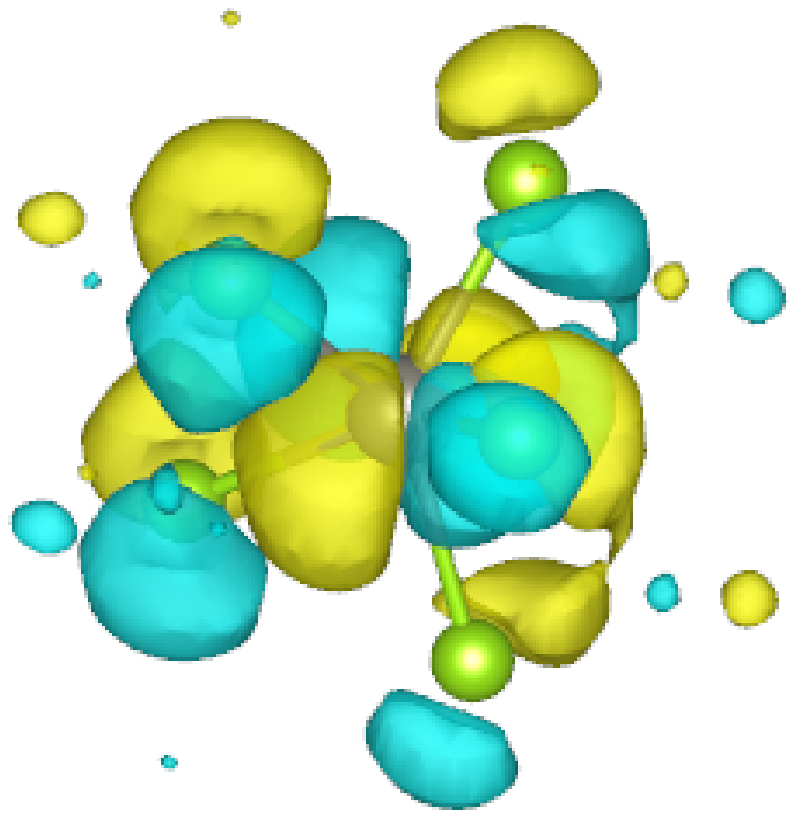}
		\caption{$d_{xy}$}
		\label{fig:dxy}
	\end{subfigure}
	~ 
	\begin{subfigure}[b]{0.3\textwidth}
		\includegraphics[width=\textwidth]{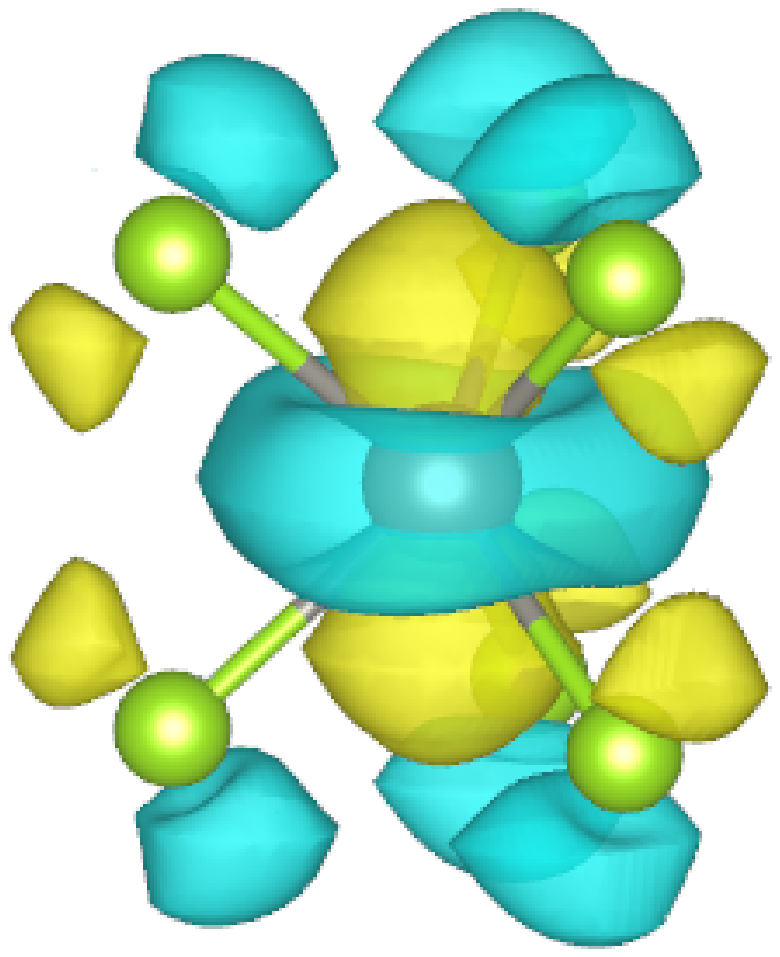}
		\caption{$d_{z^2}$}
		\label{fig:dz2}
	\end{subfigure}
	~ 
	\begin{subfigure}[b]{0.3\textwidth}
		\includegraphics[width=\textwidth]{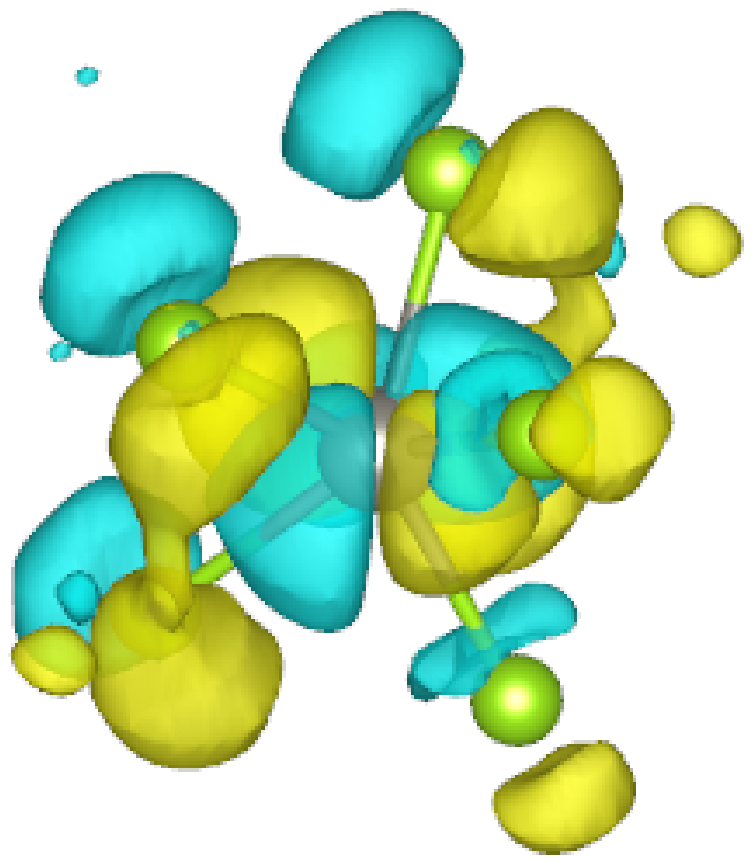}
		\caption{$d_{x^2-y^2}$}
		\label{fig:dx2-dy2}
	\end{subfigure}
	\caption{(Color online) Amplitude isosurface contours for Tungten's $d$-like Wannier functions in the three-band model. The green balls represent selenium while the gray ball represent Tungsten.}
	\label{fig2}
\end{figure}

Our goal in this work is to produce a tight-binding model that correctly 
describes the behaviour of the top of the valence band and the bottom
of the conduction band of a monolayer of WSe$_{2}$. 
Two different basis sets of Wannier functions were used to write the
one-electron Hamiltonian matrix elements.
The first one involves a similar scenario as the one presented by Liu \textit{et al.,}\cite{3band} with the calculation of the lowest conduction bands (at least around K) and the top valence band using three Wannier functions with main characters W($d_{xy}$), W($d_{z^2}$) and W($d_{x^2-y^2}$) (see Fig.~\ref{fig2} to see their spatial distribution). 
In the remaining of the manuscript this model will be referred to as the
``three-band model''.

\begin{figure}[H]
	\centering
	\begin{subfigure}[b]{0.3\textwidth}
		\includegraphics[width=\textwidth]{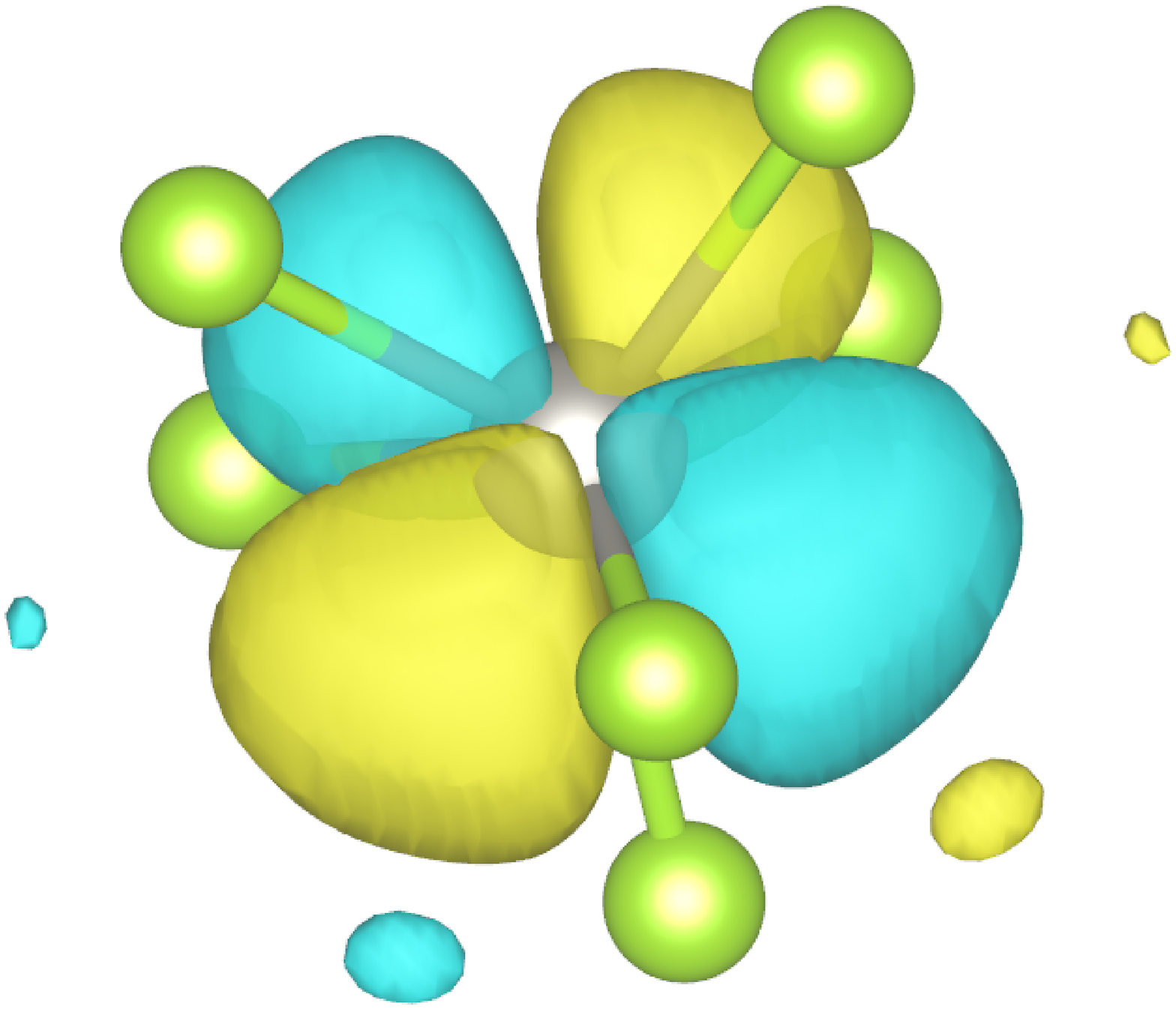}
		\caption{$d_{xy}$}
		\label{fig:dxy}
	\end{subfigure}
	~ 
	\begin{subfigure}[b]{0.3\textwidth}
		\includegraphics[width=\textwidth]{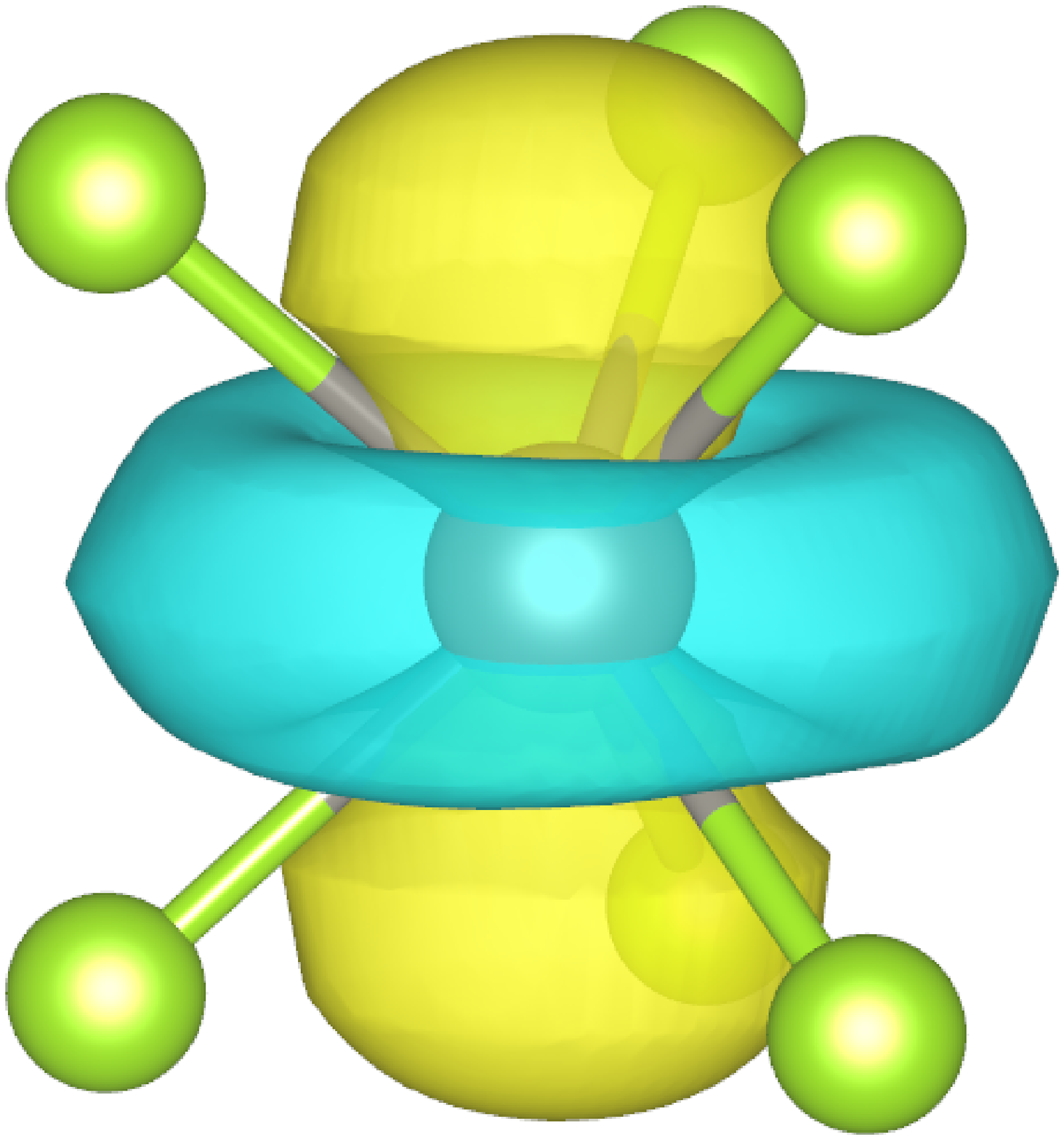}
		\caption{$d_{z^2}$}
		\label{fig:dz2}
	\end{subfigure}
	~ 
	\begin{subfigure}[b]{0.3\textwidth}
		\includegraphics[width=\textwidth]{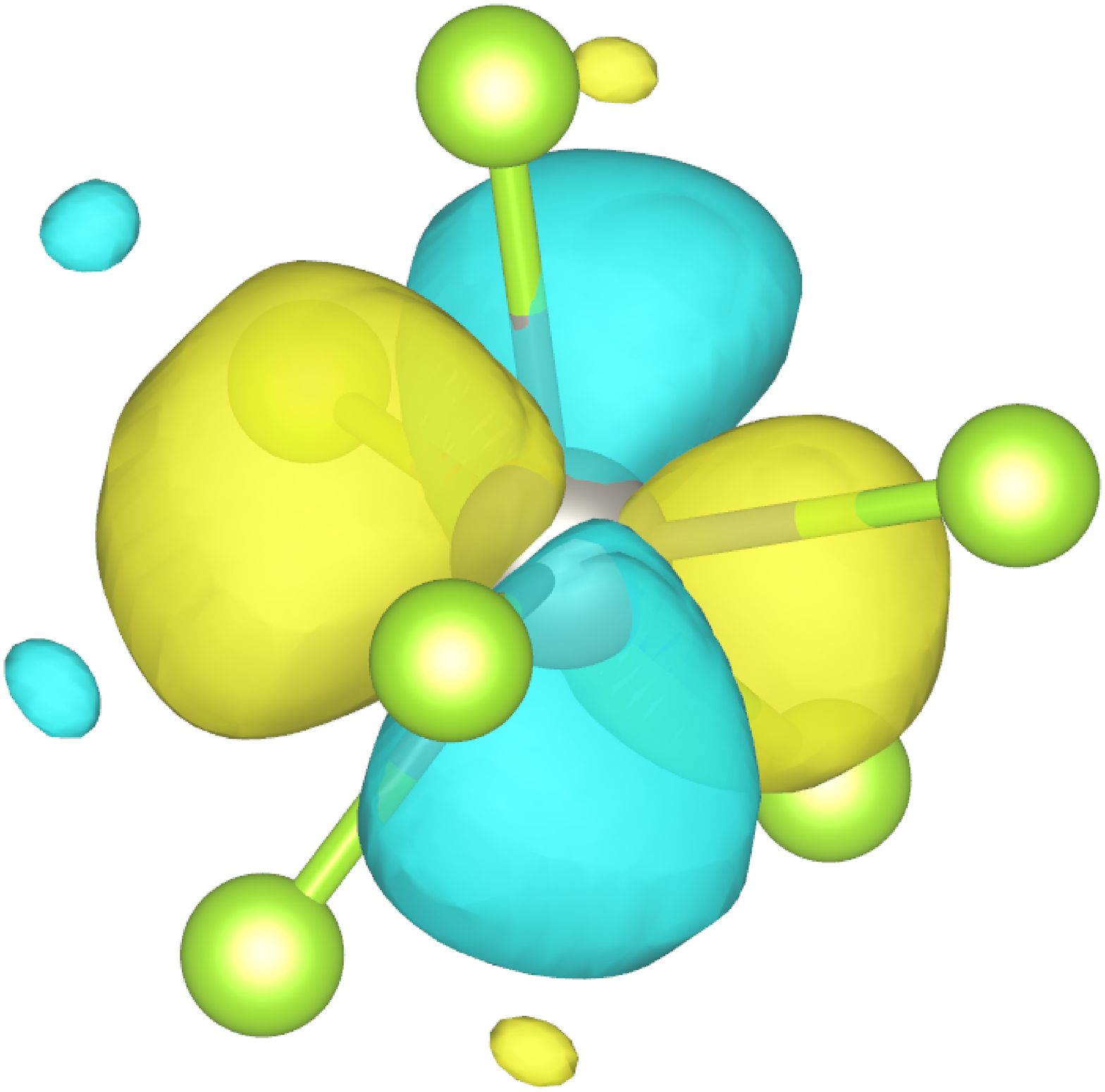}
		\caption{$d_{x^2-y^2}$}
		\label{fig:dx2-dy2}
	\end{subfigure}
	\caption{(Color online) Amplitude isosurface contours for Tungsten $d$-like Wannier functions in the nine-band model. The green balls represent selenium while the gray ball represent Tungsten.}
	\label{fig3}
\end{figure}

\begin{figure}[H]
	\centering
	\begin{subfigure}[b]{0.3\textwidth}
		\includegraphics[width=\textwidth]{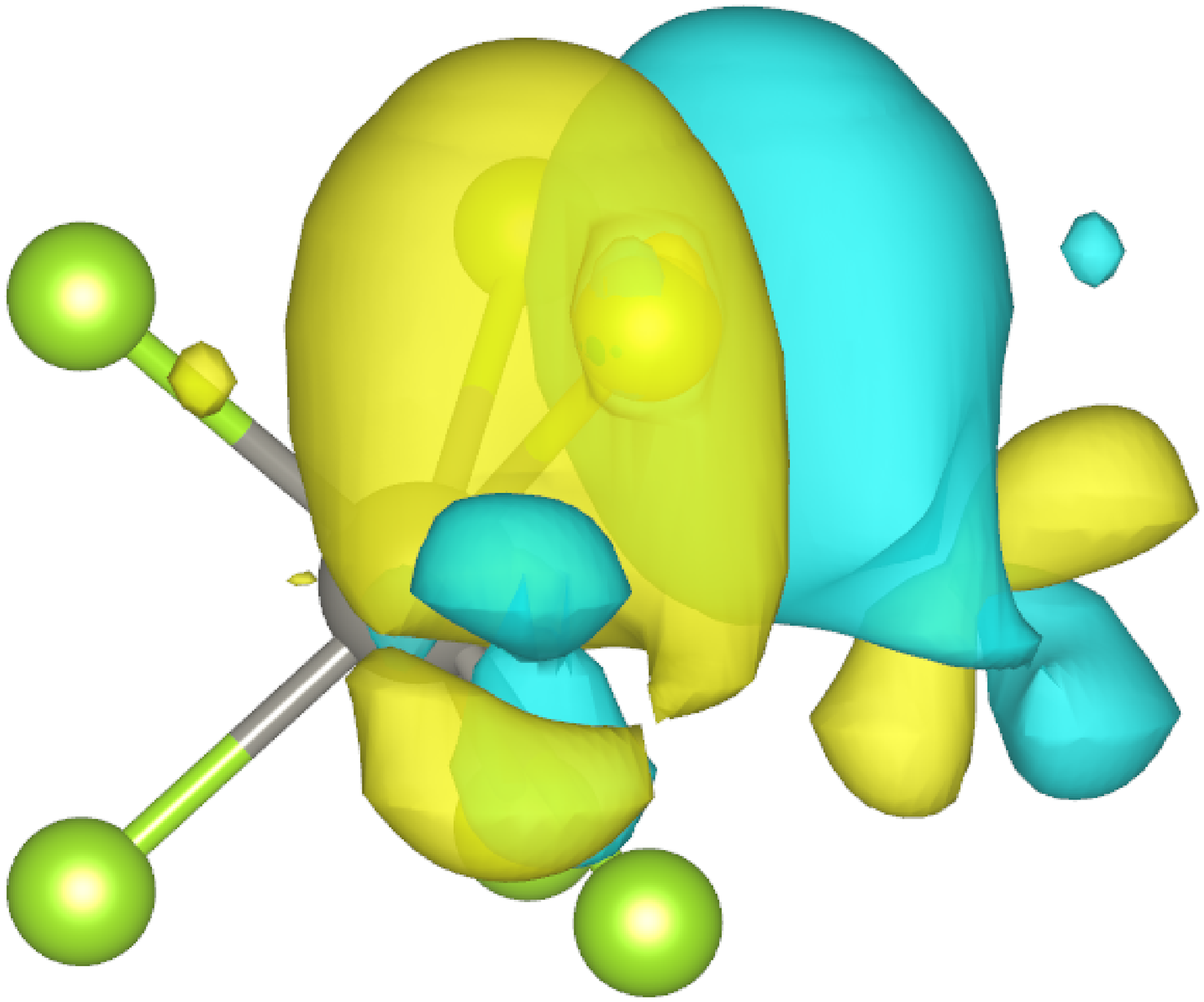}
		\caption{$p_{y}$}
		\label{fig:py}
	\end{subfigure}
	~ 
	\begin{subfigure}[b]{0.3\textwidth}
		\includegraphics[width=\textwidth]{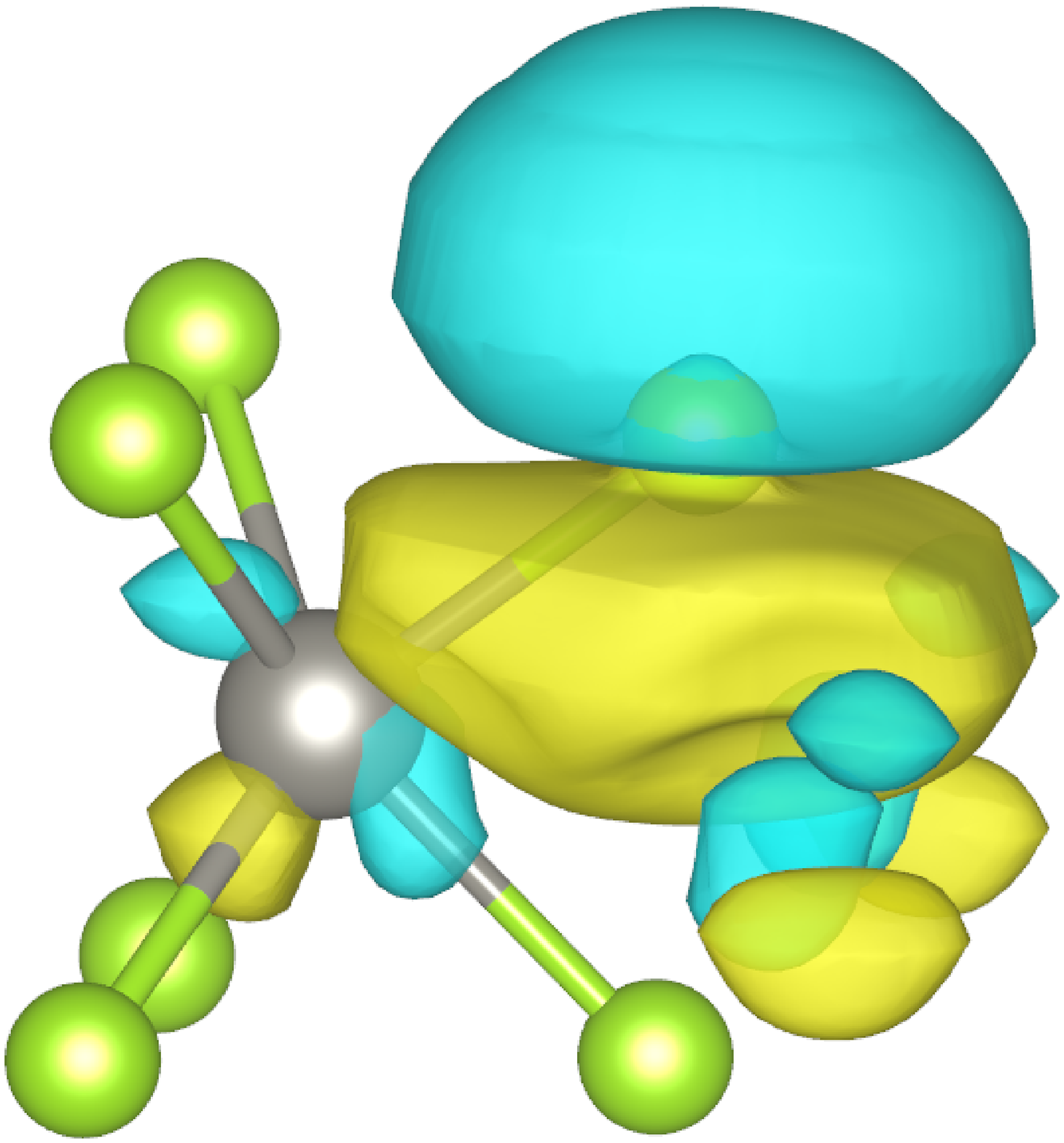}
		\caption{$p_{z}$}
		\label{fig:pz}
	\end{subfigure}
	~ 
	\begin{subfigure}[b]{0.3\textwidth}
		\includegraphics[width=\textwidth]{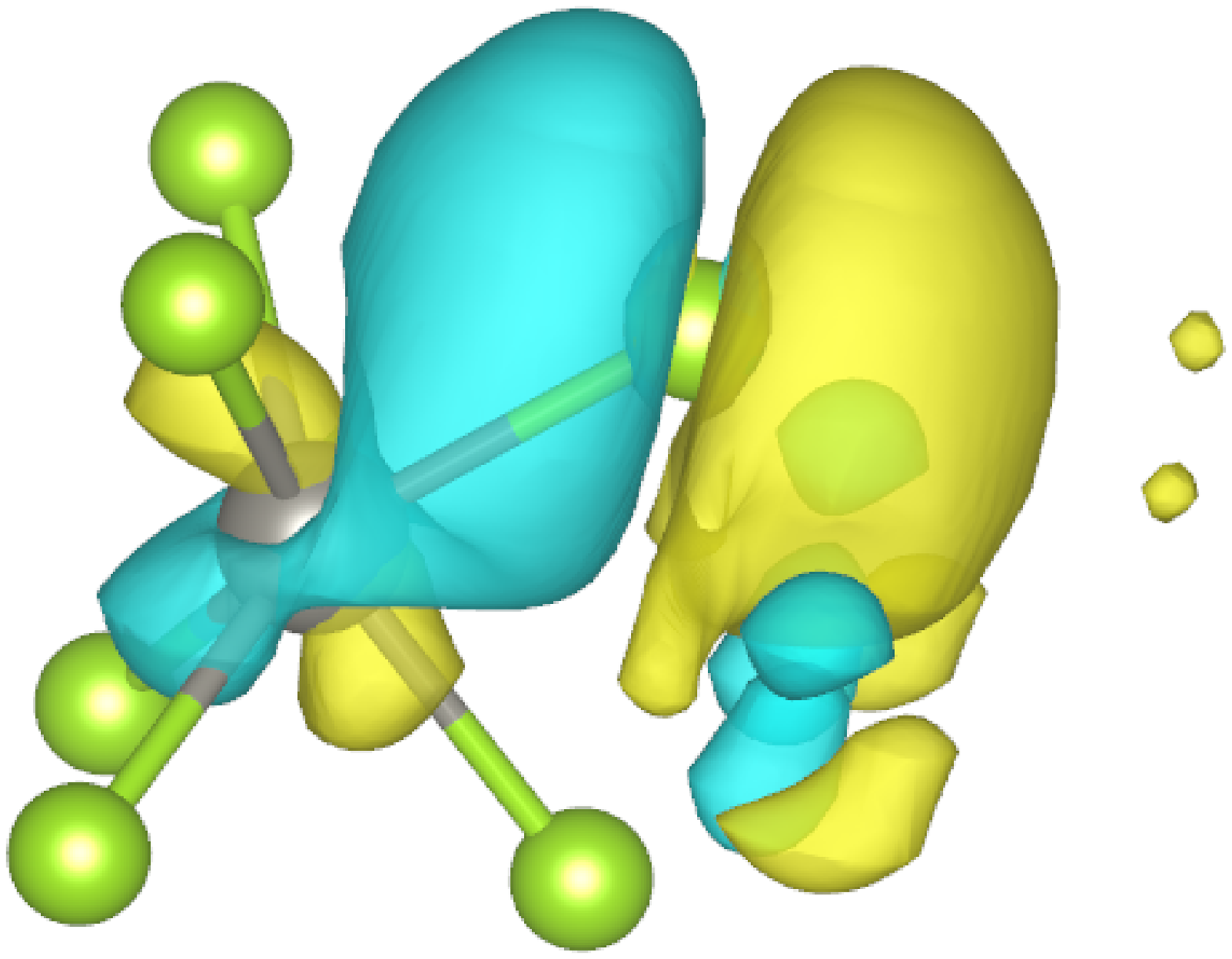}
		\caption{$p_{x}$}
		\label{fig:px}
	\end{subfigure}
	\caption{(Color online) Amplitude isosurface contours for Selenium's $p$-like Wannier functions in the nine-band model. The green balls represent selenium while the gray ball represent Tungsten.}
	\label{fig4}
\end{figure}

We shall see below how it is difficult to reproduce some of the features of 
the top of the valence bands using this basis set, mostly due to a partial
hybridization of W-$d$ orbitals with Se-$p$ orbitals.
That is the reason why we increased the number of Wannier functions in the
basis up to nine, 
where the previous model was complemented by bands with strong Se($4p$) character (see Fig.~\ref{fig3} and ~\ref{fig4}). 
This is the so-called ``nine-bands model'' below.




\subsection{Band schemes using the maximally localized Wannier functions}
\label{sec:Bandscheme}

\subsubsection{Three-band model}
\label{sec:3bandstructure}
\begin{figure*}[htb!]
	\centering
	\begin{subfigure}[b]{\textwidth}
		\includegraphics[width=0.30\linewidth,height=3.9cm]{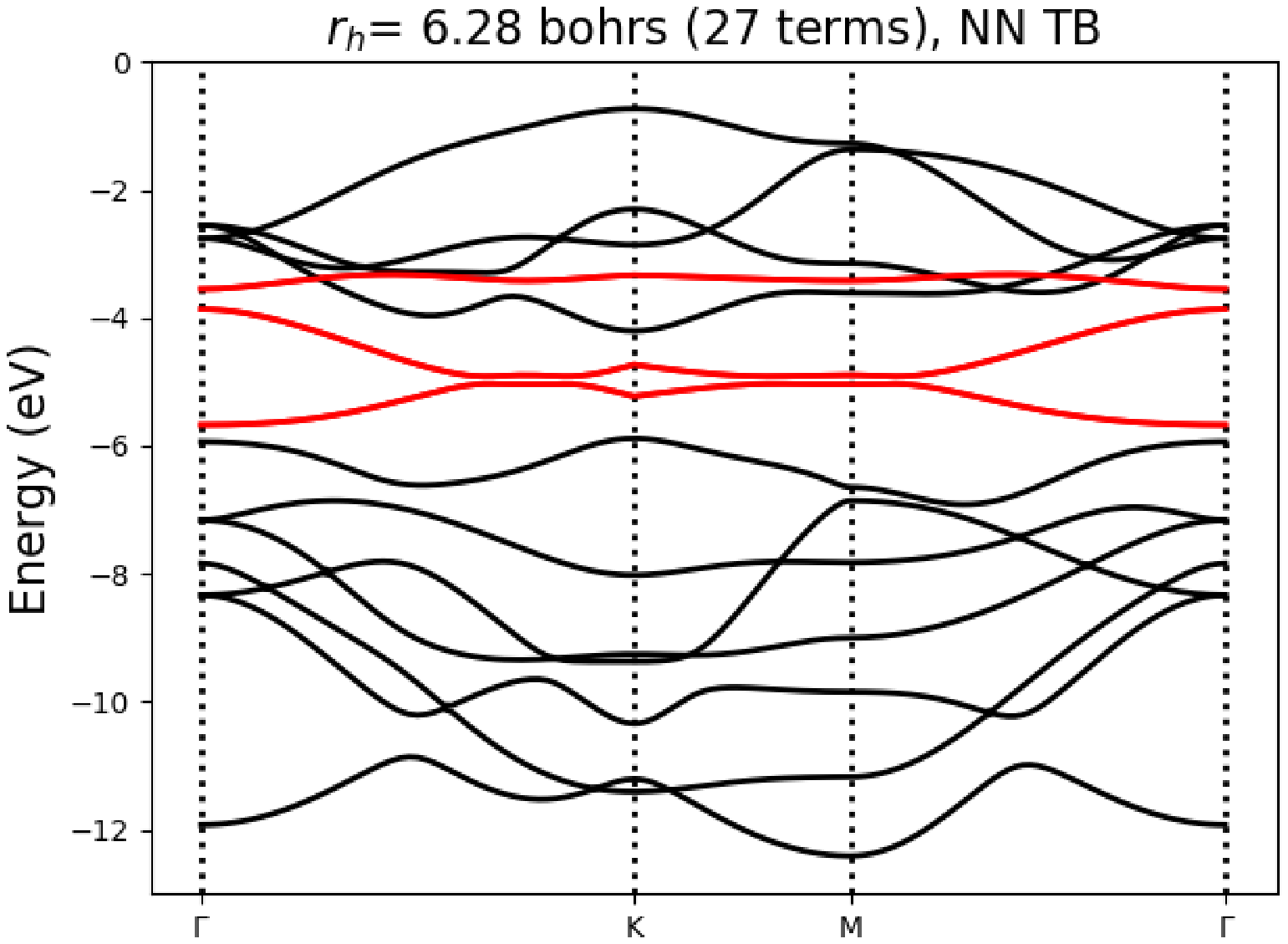}%
		\includegraphics[width=0.30\linewidth]{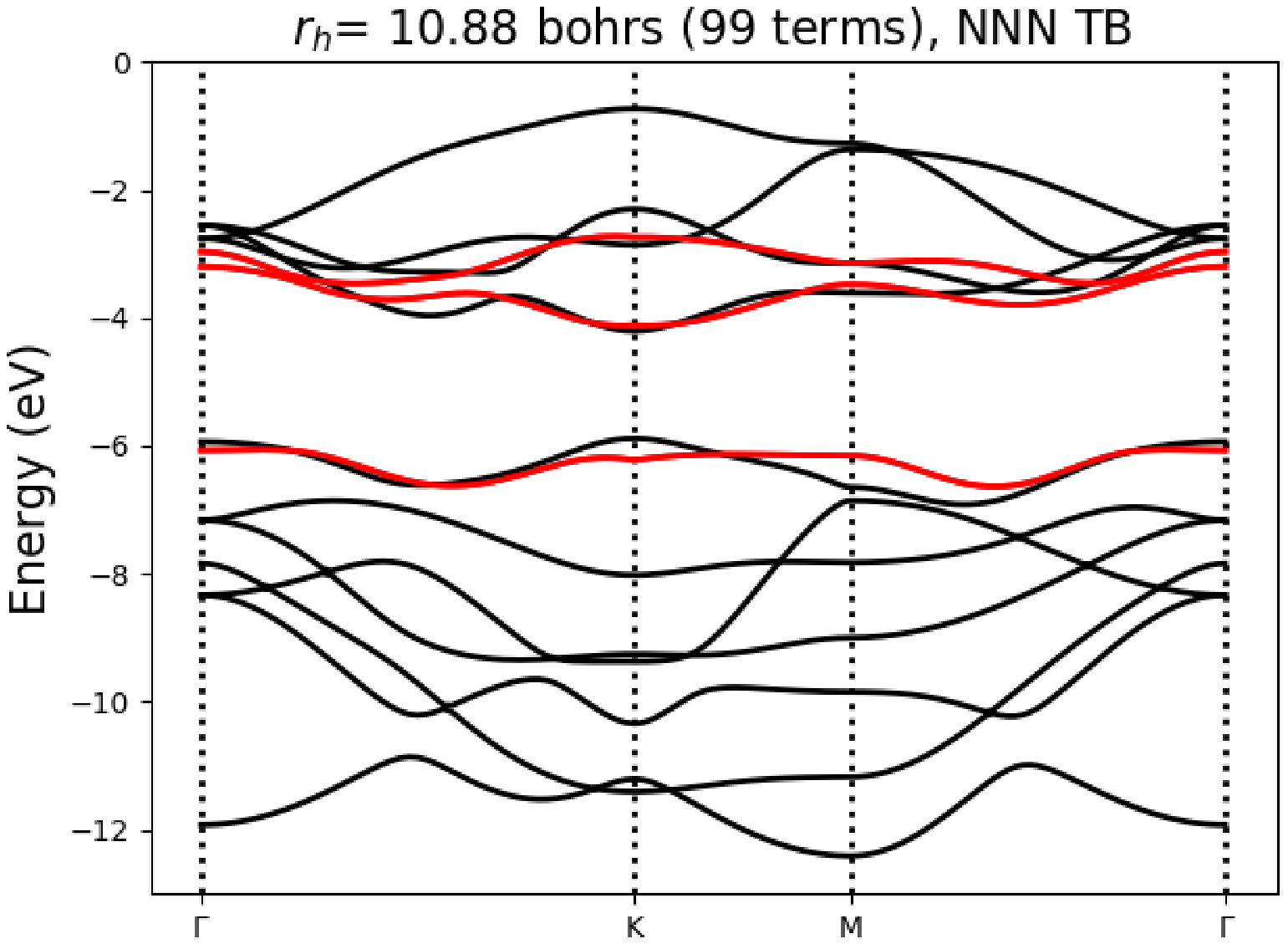}
		\includegraphics[width=0.30\linewidth]{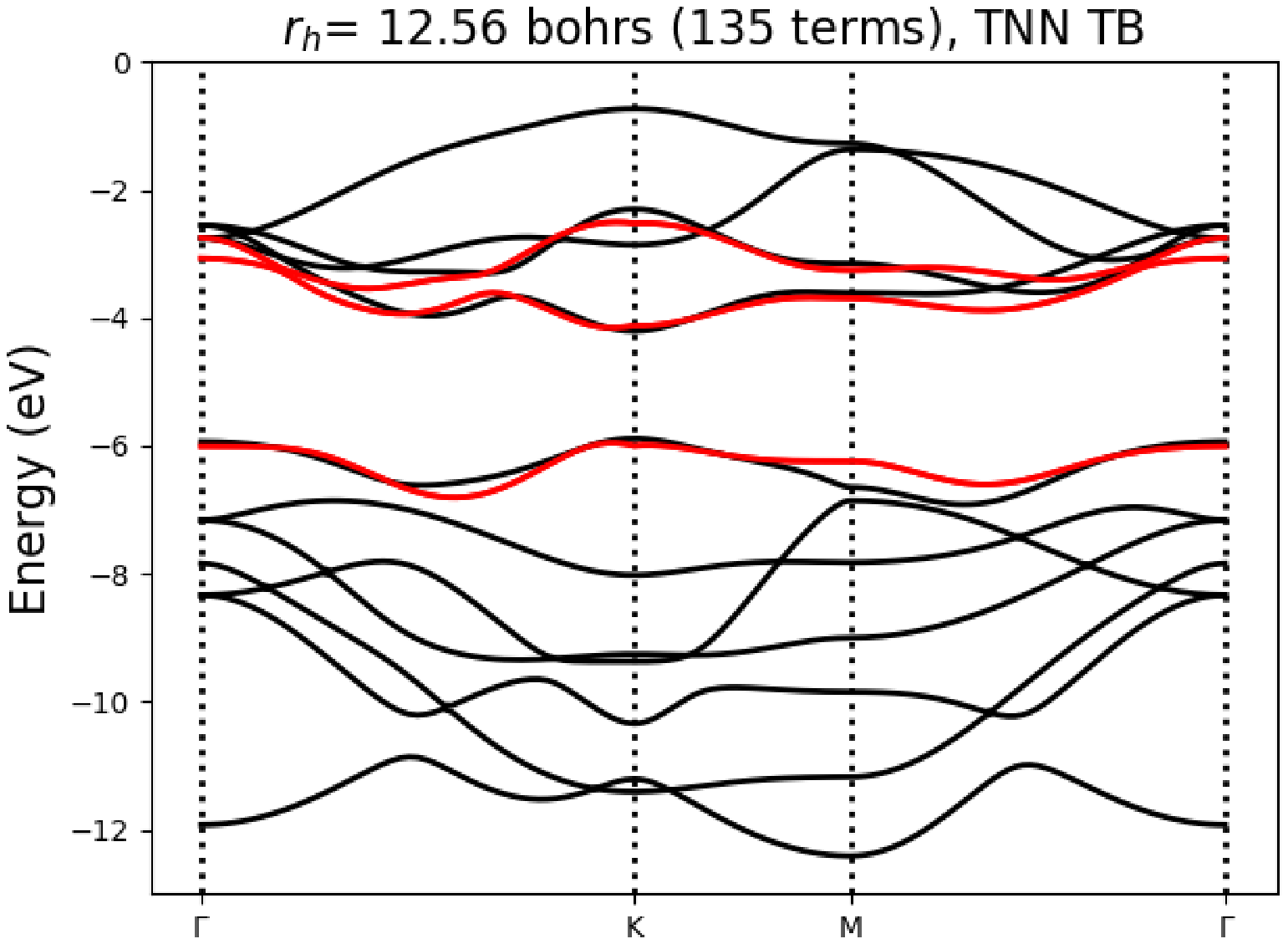}\\
		\includegraphics[width=0.30\linewidth]{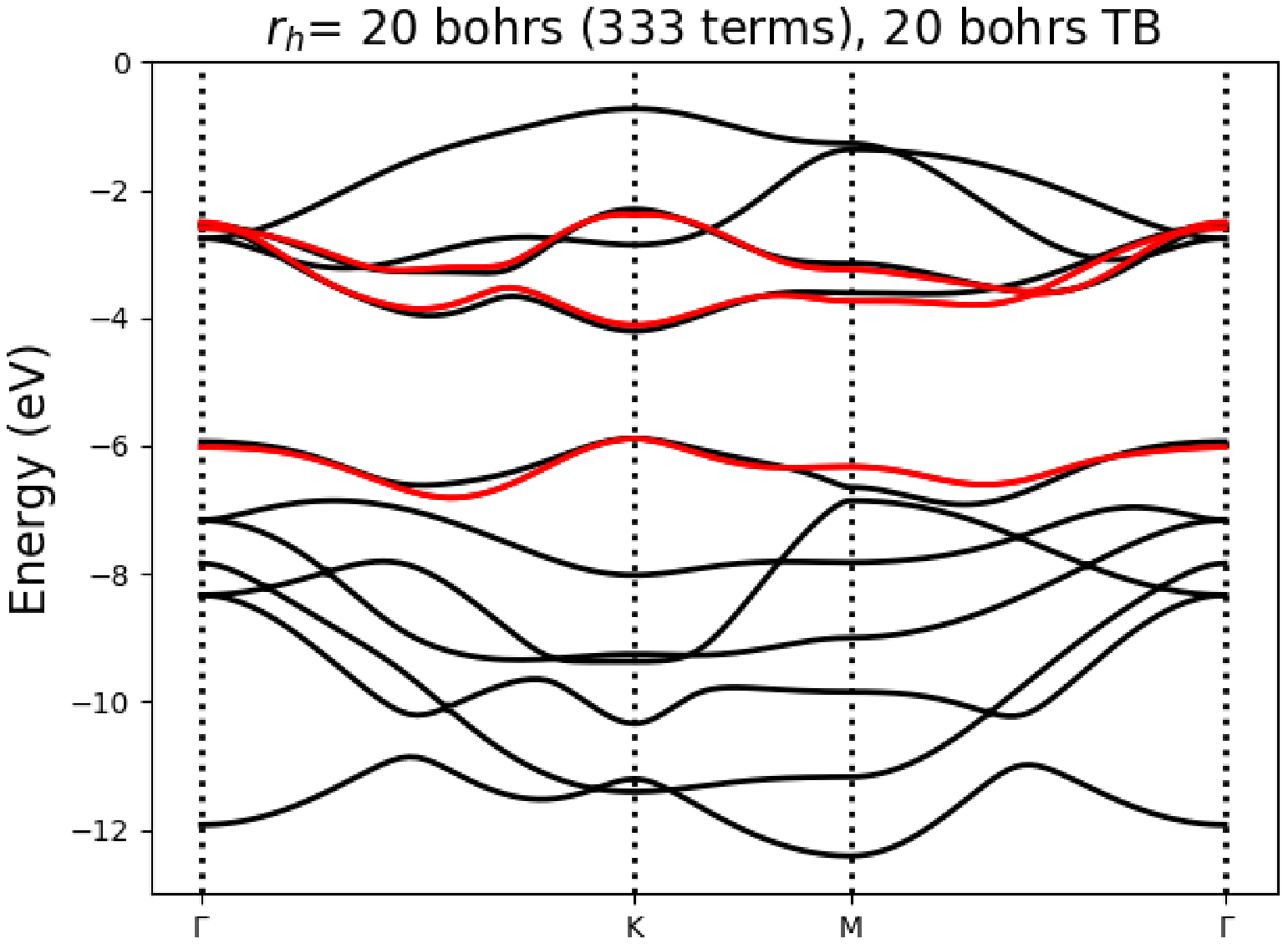}
		\includegraphics[width=0.30\linewidth]{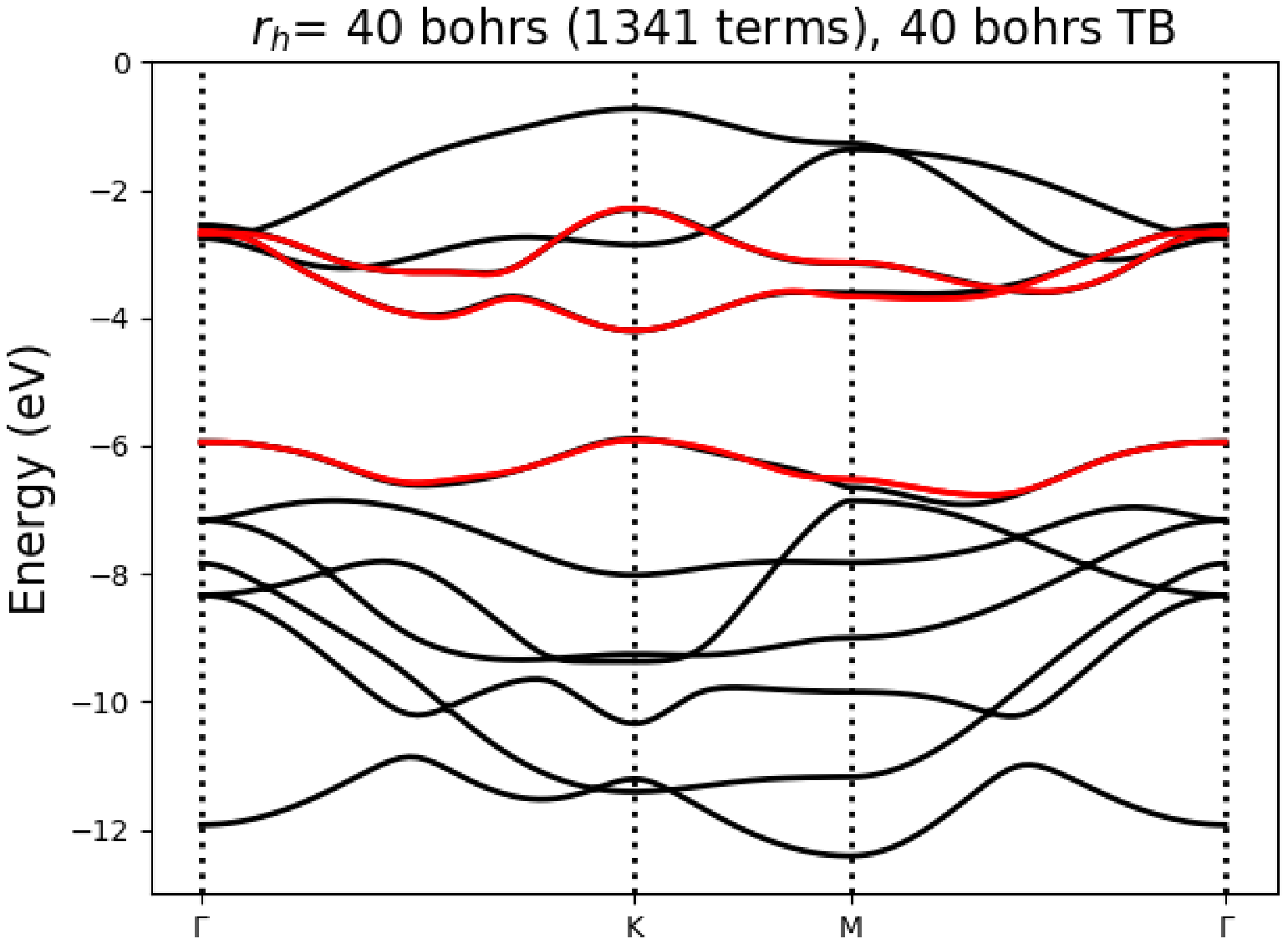}
		\includegraphics[width=0.30\linewidth]{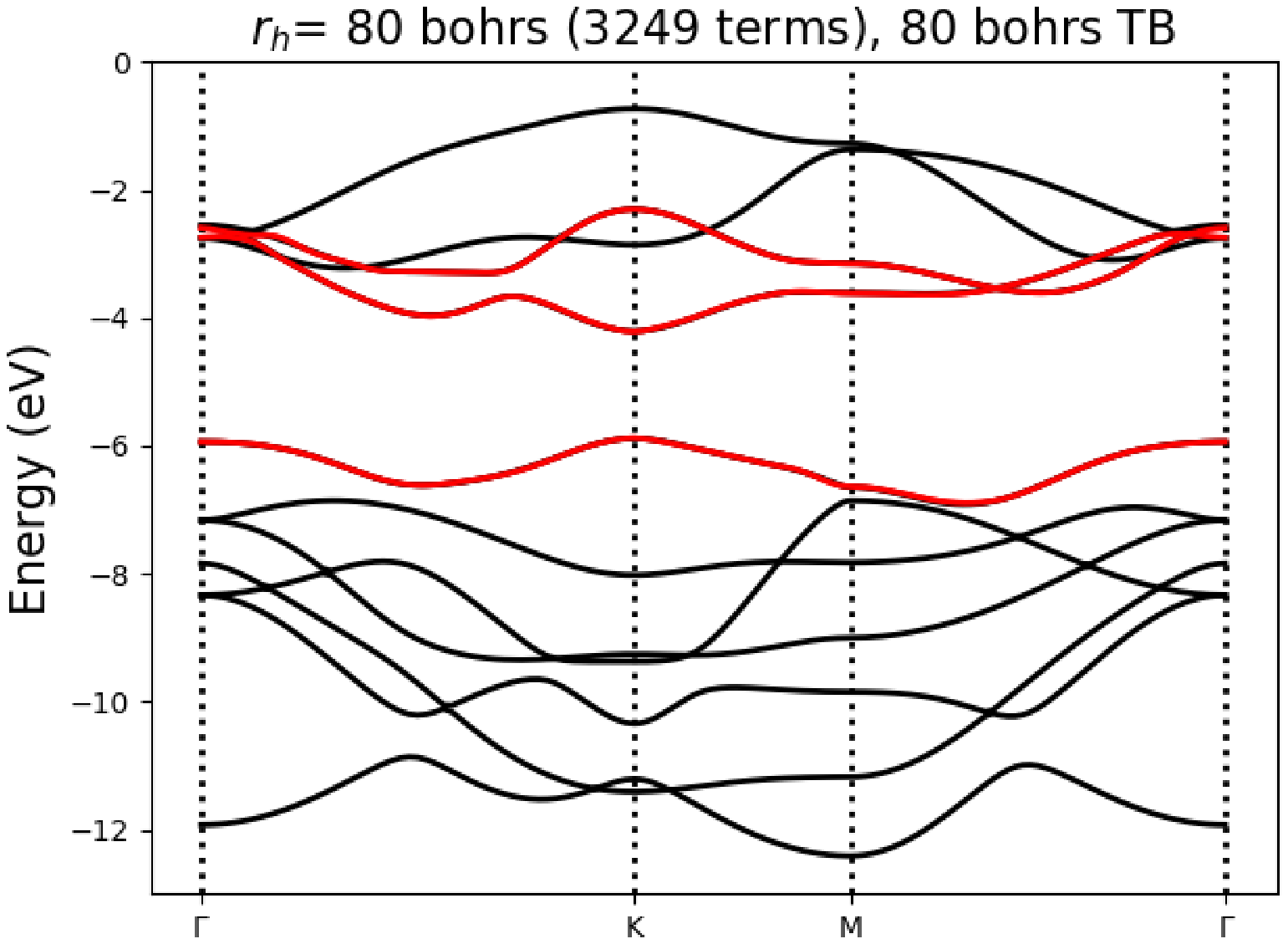}
	\end{subfigure}
	\caption{(Color online)  Band structure convergence of WSe$_2$ for various Hamiltonian cutoff, $r_h$(Where NN TB-Nearest Neighbor, NNN TB-Next Nearest Neighbor Tight-binding, TNN TB-Third Nearest Neighbor Tight-binding, and 20 Bohrs to 80 Bohrs TB). The black lines denote the bands obtained from first-principles DFT, while the results from the second-principles DFT model are represented by red lines.  At the top of each panel, we show the corresponding number of matrix elements  ($h_{ab}$) per primitive cell that were included.}
	\label{fig:3band}
\end{figure*}

Here, we compare the DFT bands with those obtained from our trimmed Wannier Hamiltonians. In the case of the three-band model the results are plotted in Fig.~\ref{fig:3band}. We chose several cutoff distances $r_h$ (see Fig.~\ref{fig:distances}) corresponding to a 6.28 Bohr (a nearest-neighbor model, NN), 10.88 Bohr (a next-nearest neighbour model, NNN), 12.56 Bohrs (a third neighbor model, TNN) and then at regular intervals, $r_h=$20, 40, 80 Bohrs. The first three $r_h$ values (NN, NNN and TNN) allows us direct comparison with the work of Liu and co-workers\cite{3band} where they fitted the DFT bands to analytic NN, NNN and TNN tight-binding models. 

The first thing we notice is that, while the band-fitted NN model in Ref.~\onlinecite{3band} was able to reproduce important segments of the band diagram particularly the top of the valence band and the bottom of the conduction one, when one explicitly constructs the Wannier basis functions for the tight-binding this does not happen (see Fig.~\ref{fig:3band}a) and the model is far from qualitatively reproducing the bands at all. This calls into question the possibility of using a NN model to perform simulations on WSe$_2$ like optical or transport  beyond a basic reconstruction of stationary state-bands. We observe that bands for $r_h$=10.88 (NNN model) and $r_h$=12.56 Bohr (TNN model) are qualitatively correct although numerically they are still quite far apart from reproducing the DFT bands. This is again, in strong contrast with the results of Liu \textit{et al.,}\cite{3band} where their analytical TB model does a very good job of reproducing qualitatively and quantitatively the DFT bands. 
Certainly, the effect of farther neighbour shells can be effectively 
considered in the parameters of the models with only NN.
However, since these results are not supported by Wannier function 
calculations (and all tight-binding models assume the existence of 
Wannier-like functions), it would be important to check whether those models
provide sensible results for magnitudes beyond the reproduction of 
band-dispersion curves they were fitted to.

For $r_h=$20.0 Bohr (equivalent to a nearest 4$^{th}$ neighbor model, see Fig.~\ref{fig:distances}) we observe that the model finally can reproduce quantitatively the bands except around M-point, particularly at the top of the valence band. These variations are almost completely corrected for $r_h$=40.0 Bohr in the conduction band and at $r_h$=80.0 Bohr for the valence band. The problem with the valence band close to M-point is the close vicinity of another band with strong Se($4p$) character. Both bands have an avoided crossing at that point, are strongly entangled and it is difficult to construct an accurate three-band model at that point. 

When we observe the Wannier functions corresponding to this model (see Fig.~\ref{fig2}) we see that, indeed, the bands have strong W($d_{z^2}$), W($d_{xy}$) and W($d_{x^2-y^2}$) character. However, we see that these functions are strongly hybridized with Se($4p$) orbitals and are not optimally compact, explaining the need to use the large cutoff values discussed above. 

\subsubsection{Nine-band model}
\label{sec:9bandstructure}

\begin{figure*}[htb!]
	{\centering
	\begin{subfigure}[b]{\textwidth}
		\includegraphics[width=0.33\linewidth]{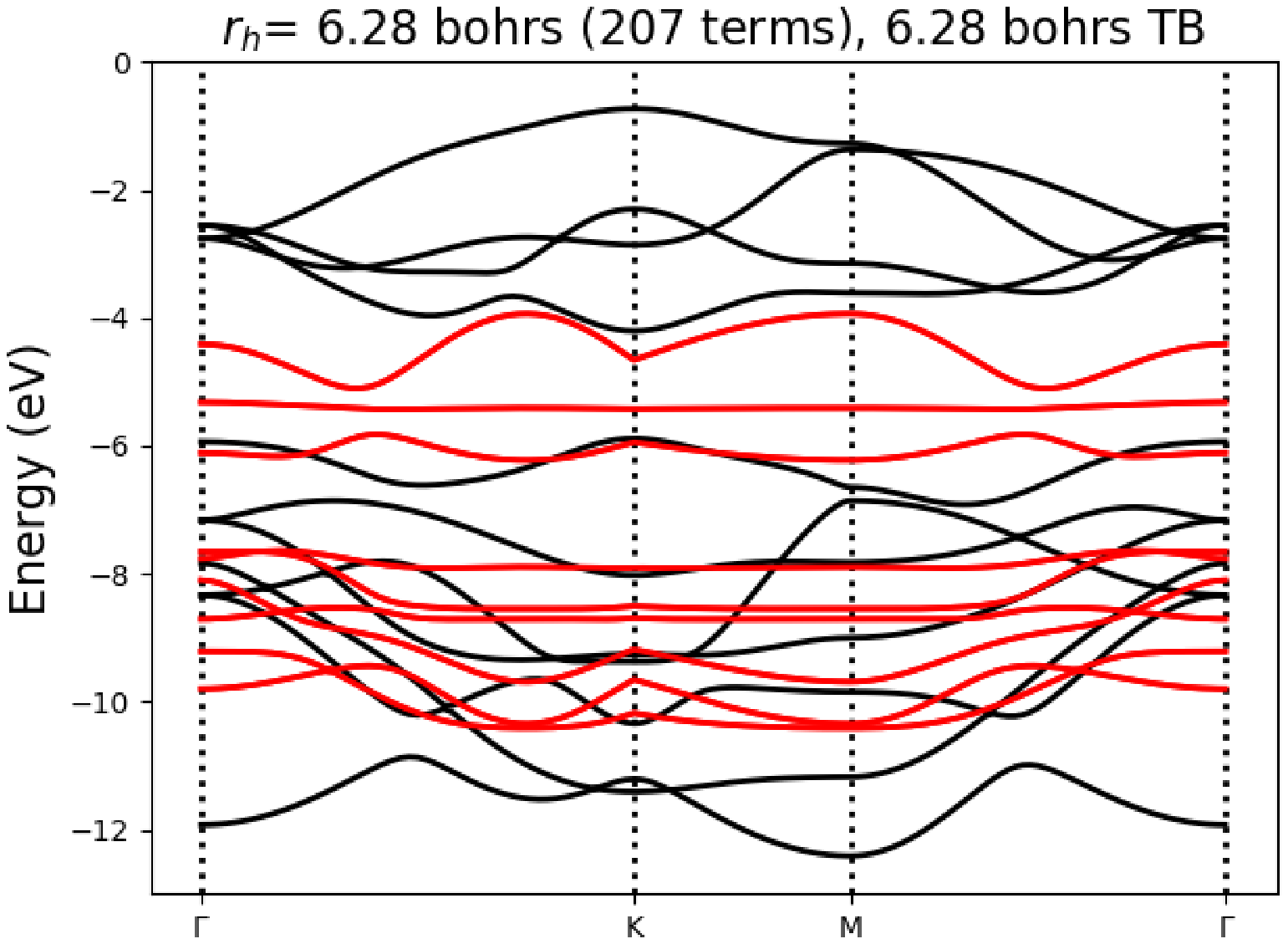}%
		\includegraphics[width=0.33\linewidth]{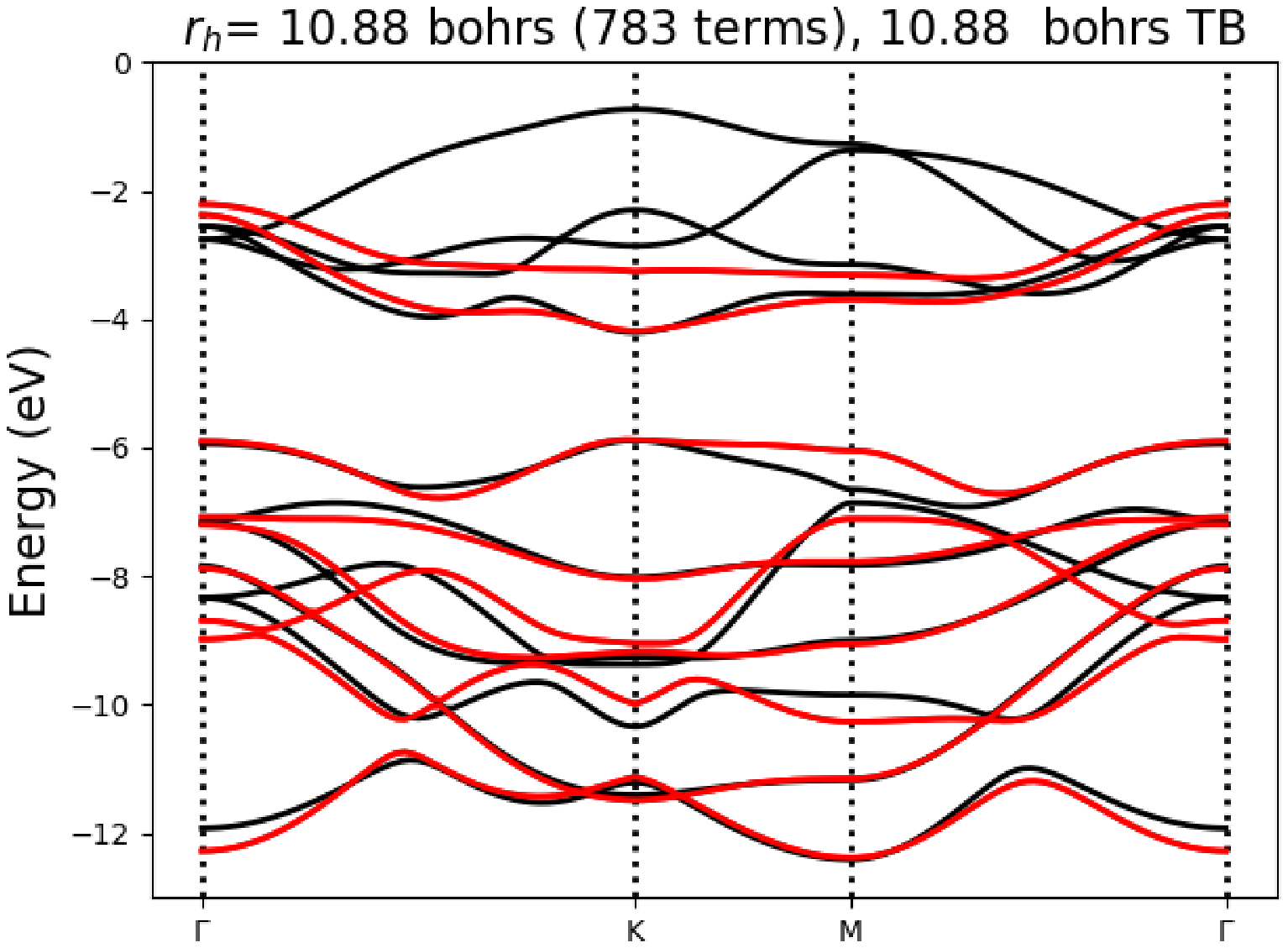}
		\includegraphics[width=0.33\linewidth]{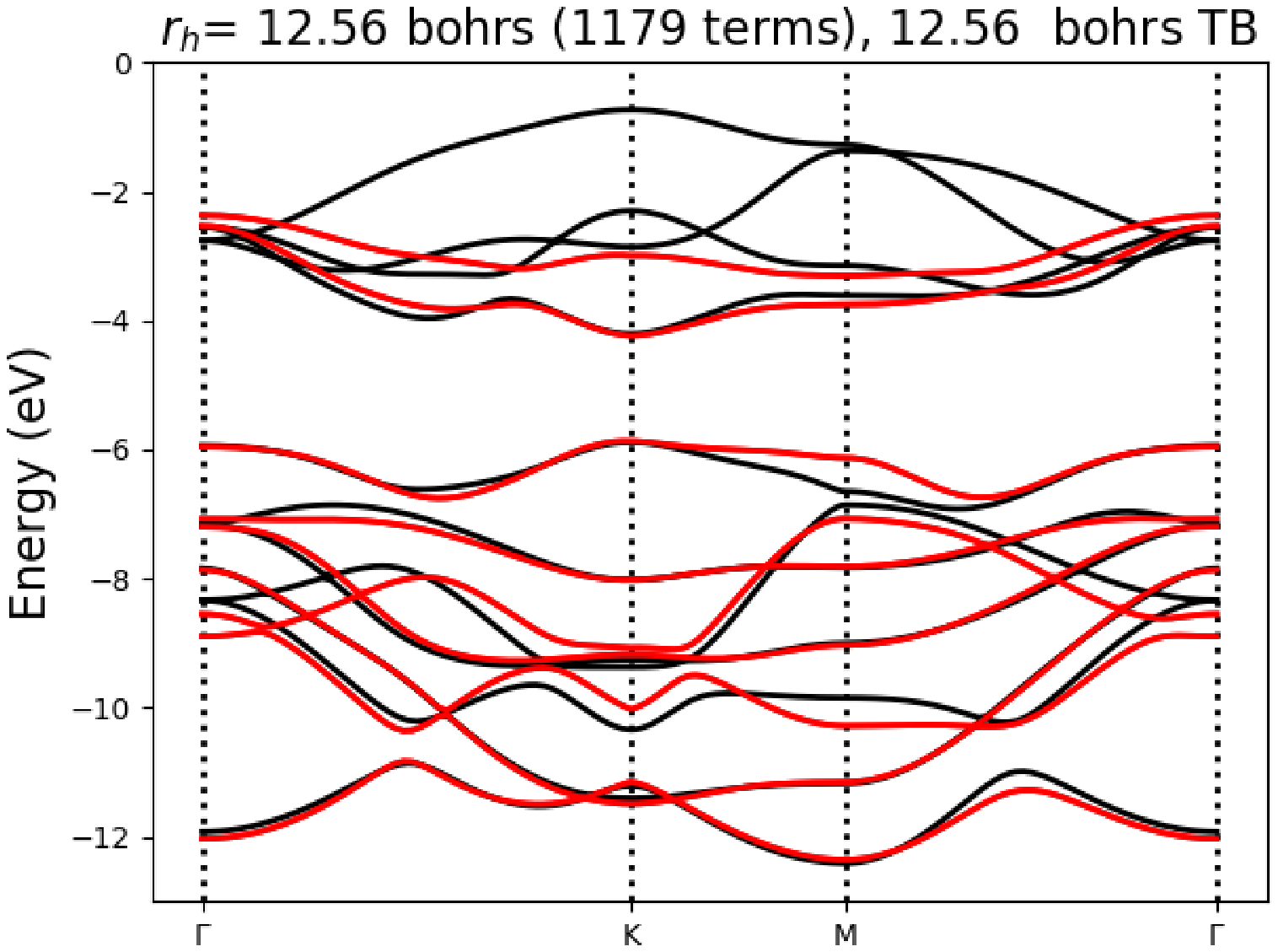}
	\end{subfigure}
	\vskip\baselineskip
	\begin{subfigure}[b]{\textwidth}
		\includegraphics[width=0.33\linewidth]{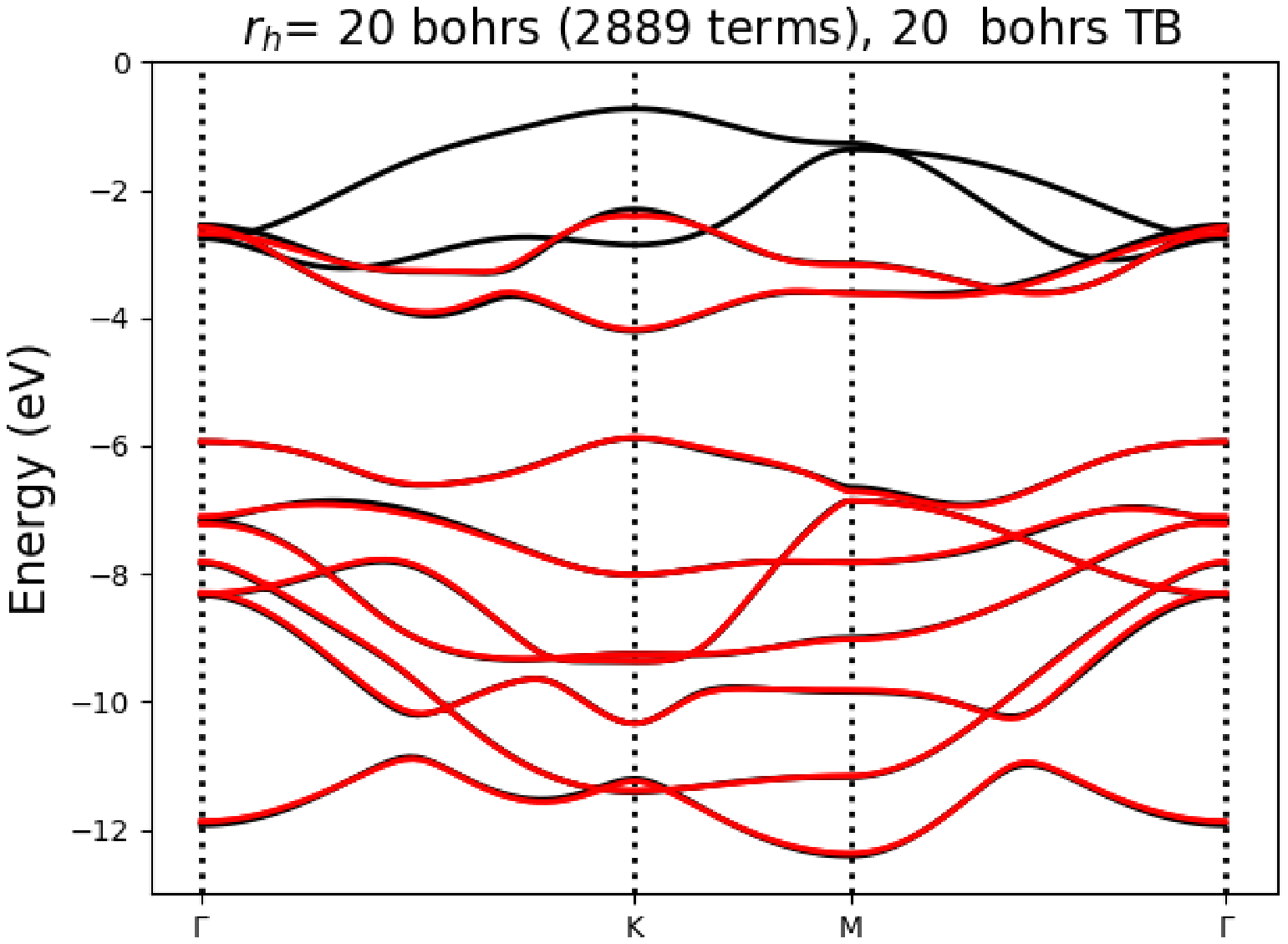}%
		\hfill
		\includegraphics[width=0.33\linewidth]{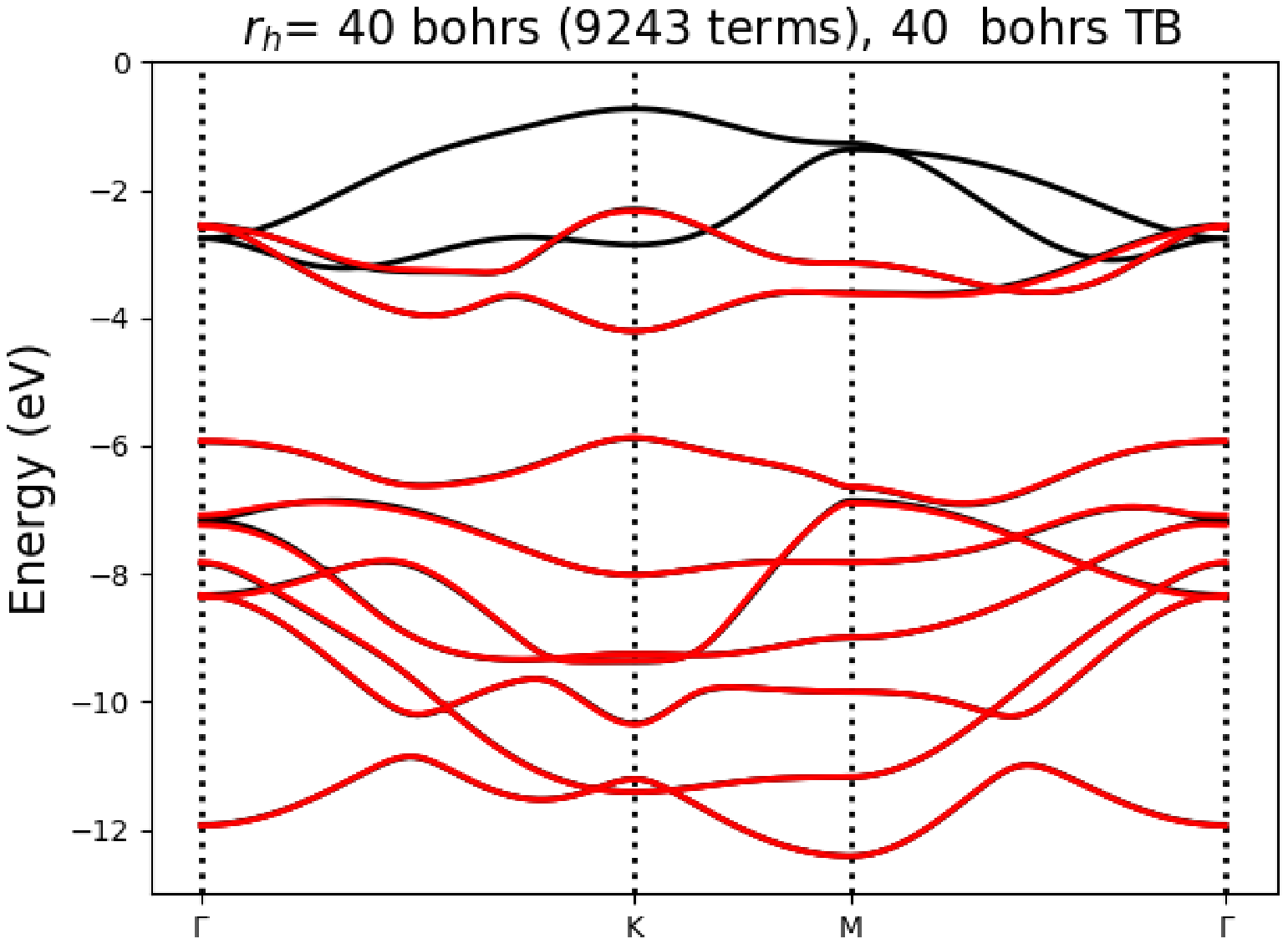}
			\hfill
		\includegraphics[width=0.33\linewidth]{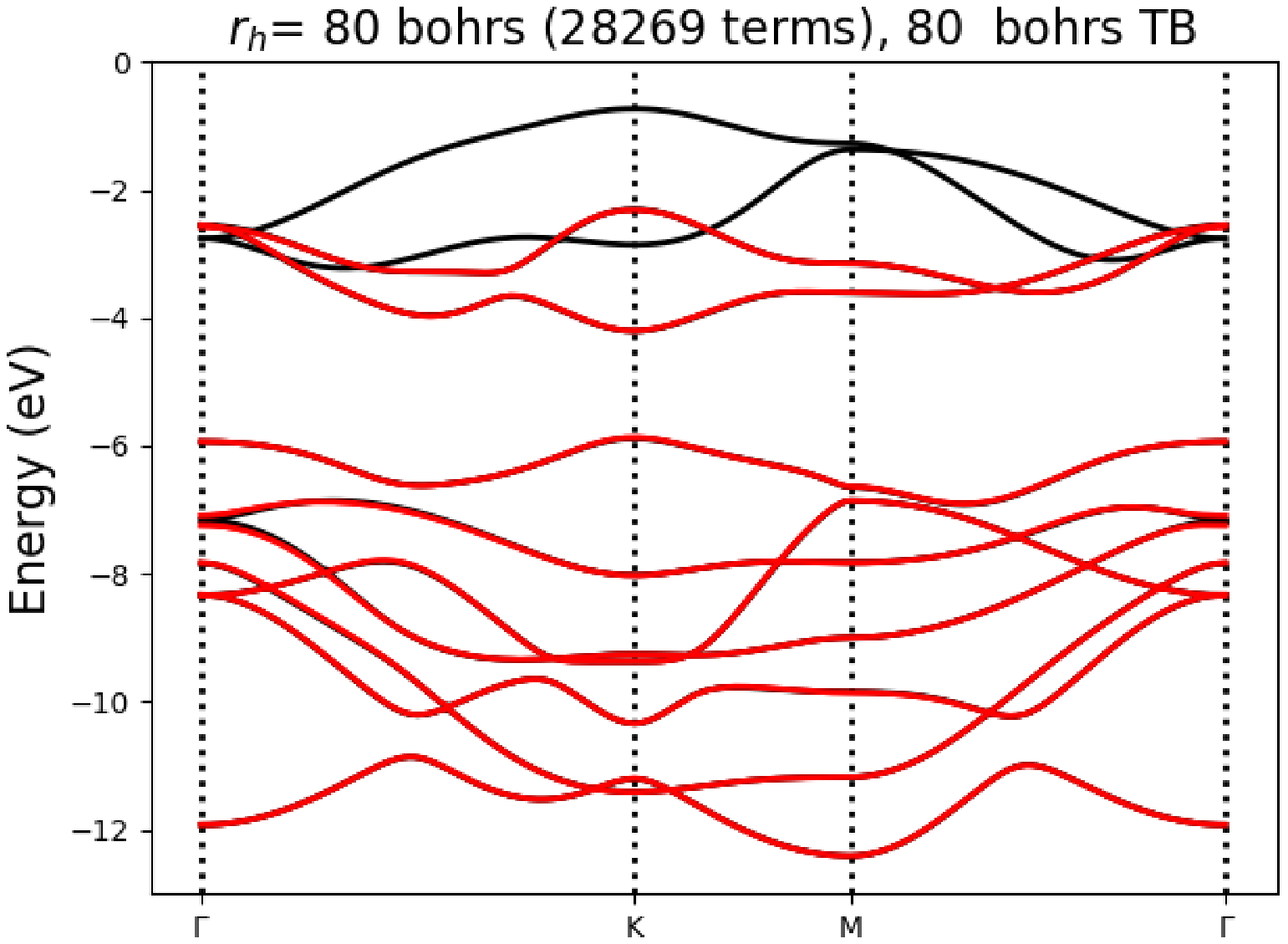}
	\end{subfigure}
        }
	\caption{(Color online) Representation of the band structure of WSe$_2$ for various Hamiltonian cuttoff, $r_h$. The black lines denote the bands obtained from first-principles DFT, while the results from the second-principles DFT model are represented by red lines.  At the top of each panel, we show the corresponding number of matrix elements ($h_{ab}$) per primitive cell that were included.}
	\label{fig1}
\end{figure*}

In Fig.~\ref{fig1} we compare the full first-principles band structure for WSe$_2$ with that obtained from the nine-band model using different cutoff radii. As it occurred in the three-band model the NN models yields bands that are qualitatively wrong, particularly the conduction ones. Upon increasing the range of interactions to NNN or TNN the valence and conduction bands are much better reproduced although they still present significant discrepancies, particularly in the lowest conduction bands. However, going to $r_h$=20.0 Bohr leads to bands that match the DFT ones all over the first Brillouin zone. When we compare the band gap obtained in this model (1.50 eV) with that given by Swastibrata and Abhishek\cite{goodref} we find they are in excellent agreement. 

Compared with the three-band model we observe a relatively quick convergence with $r_h$ for two reasons: (i) the Wannier orbitals have in this case a clear-cut Tungsten-mainly and Selenium-mainly character (see Figs.~\ref{fig3}, \ref{fig4}) and are thus more localized, (ii) the inclusion of W and Se bands allows to efficiently describe the avoided crossing occurring around the top of the valence band in the M-point. This is thus a sensible choice to base the construction of a full second-principles model for WSe$_2$ where the inclusion of lattice distortions would produce more problems with the band mixing at the points of the reciprocal space where bands of different character are in close vicinity.

\section{CONCLUSION}
\label{sec:conclusion}
In this paper, we have developed Wannier-based tight-binding models for a monolayer of WSe$_2$ using three-band and nine-band models to reproduce the valence and lower conduction bands. For the three-band model the Wannier functions display strong W($d_{z^2}$), W($d_{xy}$), W($d_{x^2-y^2}$) character and, in order to reproduce the bands over the whole Brillouin zone, a large cutoff radius, $r_h$, must be chosen. This result contrasts with the semi-analytic tight-binding models of Liu \textit{et al.,}~\cite{3band} where a simple NN neighbor model is used to reproduce the top valence and bottom conduction band and a TNN model is used to obtain a good fitting of the dispersion of the bands over the whole reciprocal space. Given that the methods used by Liu \textit{et al.,} are not based on the calculation of a real localized function to create their matrix elements we believe that one should be careful with their use beyond reproducing the shape of the bands. 

We also find that a nine-band model with a 20 Bohr cutoff is able to nicely reproduce the band diagram. This model is able to efficiently describe several focal points of the electronic structure of the system like a band crossing taking place on the top of the valence band between M and $\Gamma$ points and we believe it is an excellent starting point to further build a full second-principles model for WSe$_2$ including lattice, electron-lattice, electron-electron and spin-orbit effects in order to simulate the many exciting properties in the WSe$_2$  monolayer. We would finally like to add that construction of tight-binding models allows to easily check the effect of increasing the basis of localized functions and the effect on the effective range of interaction. This ability of systematic model checking and improvement is a large advantage with respect to usual ways to construct tight-binding models.

\section*{acknowledgement} 
\label{sec:Acknowledgement} 
J.S. thanks the University of Cantabria for the scholarship funded by the Vice-rectorate for Internationalization and the Theoretical Condensed Matter Group. J. S. also acknowledges the support of the African School on Electronic Structure Methods and Applications (ASESMA). J.J. and P.G.-F. acknowledge financial support from the Spanish Ministry of Economy and Competitiveness through the MINECO Grant No. FIS2015-64886-394-C5-2-P, and the Spanish Ministry of Science, Innovation and Universities through the grant No. PGC2018-096955-B-C41. G.S.M acknowledges the support of the Kenya Education Network through CMMS mini-grant 2019/2020. The authors also gratefully acknowledge the computer resources, technical expertise, and assistance provided by the Centre for High Performance Computing (CHPC), Cape Town, South Africa. 

\section*{AUTHOR CONTRIBUTIONS} 
\label{sec:contributions}
The research problem was designed by J.S and P.G.-F. All the calculations were done by J.S. Data analysis was done by all the authors. The manuscript was written by J.S and G.S.M and revised by G.A, P.G.-F, J.J. All the authors approved the manuscript for submission.

\end{document}